\documentclass{fairmeta}
\usepackage{graphicx}
\usepackage{natbib}
\usepackage{caption}
\usepackage{algorithm}
\usepackage{algorithmic}
\usepackage{booktabs}
\usepackage{amsmath}
\usepackage{amssymb}
\usepackage{multirow}
\usepackage{placeins}
\usepackage{comment}
\usepackage{tikz}
\usetikzlibrary{arrows.meta,positioning}

\newcommand{\method}{GLocFM}
\newcommand{\vect}[1]{\mathbf{#1}}
\newcommand{\R}{\mathbb{R}}

\newcommand{\nrx}{N_{\!R}}

\title{\method{}: A Geometry-Aware Foundation Model for 3D Indoor Wireless Localization}
\author[\dagger 1]{Chenghong Bian}
\author[\dagger 1]{Chaozheng Wen}
\author[1]{Hongze Chen}
\author[*1]{Jun Zhang}
\affiliation[1]{Hong Kong University of Science and Technology}
\contribution{\textsuperscript{$\dagger$}Equal Contributions\quad\textsuperscript{*}Corresponding Author}
\metadata[Correspondence to]{Jun Zhang (\email{eejzhang@ust.hk})}

\begin{document}

\abstract{
Learning-based wireless localizers often fail to utilize geometric information about the propagation environment, limiting their ability to exploit non-line-of-sight (NLoS) propagation and generalize across scenes.
To bridge this gap, we propose \textbf{\method{}}, a \textbf{\underline{G}}eometry-aware \textbf{\underline{Loc}}alization \textbf{\underline{F}}oundation \textbf{\underline{M}}odel, which jointly exploits WiFi measurements and scene geometry represented as a 3D point cloud.
%To mitigate the gap, we present \method{}, a Geometry-aware Localization Foundational Model that jointly leverages WiFi signals and scene geometry represented by a 3D point cloud. 
We formulate localization as a maximum-likelihood (ML) estimation problem, where the goal is to find a transmitter position that maximizes the likelihood of the wireless observations conditioned on the scene geometry. The likelihood of a candidate transmitter position is calculated by a learned scoring function that matches the observed delay--angle-of-arrival (AoA) spectrum against the spectrum predicted for that candidate. A hierarchical scene encoder extracts propagation-relevant features to produce geometric priors for LoS and one-bounce reflection paths. For scenarios with imperfect synchronization, we further introduce a time-of-flight (ToF)-robust \method{} model to handle unknown ToF offsets.
\method{} is trained on a multi-modal synthetic indoor localization dataset
comprising 221 diverse scenes whose associated wireless signals are generated
using Sionna RT.
On both synthetic and the NeRF$^{2}$ dataset based on real measurements, \method{} reduces mean 3D
localization error relative to one of the state-of-the-art localization baselines by  49.5\% and 48.8\%, respectively.
Ablations across different number of receiver, bandwidths, and array sizes further
demonstrate the effectiveness and robustness of the proposed framework.

}

\maketitle

\section{Introduction}
%Indoor localization supports robot navigation, asset tracking, and context-aware services where satellite positioning is unreliable [please cite]. 
Accurate indoor localization is a key enabler for autonomous robot navigation, real-time tracking of equipment and inventory, and services in environments where global positioning system (GPS) signals are unavailable or severely attenuated \cite{zafari2019indoorlocalizationsurvey}.
{Literatures \cite{liu2007indoorpositioningsurvey,zafari2019indoorlocalizationsurvey} identify NLoS blockage, multipath propagation, material-dependent attenuation as persistent challenges for practical indoor localization, as they make the mapping from wireless observations to transmitter positions ambiguous and scene-dependent.}
By properly utilizing geometric information, we can identify potential multipath components and thereby improve our understanding of the wireless observations. Moreover, by exploiting geometric priors, it is possible to determine whether a position could contain a transmitter, motivating efficient localization algorithms that use this information to improve performance.

\begin{figure}[!t]
    \centering
    \IfFileExists{figures/glocfm_overview.pdf}{
        \input{figures/system_ml_pipeline_overlay}
    }{
        \PackageError{main}{Missing required figure 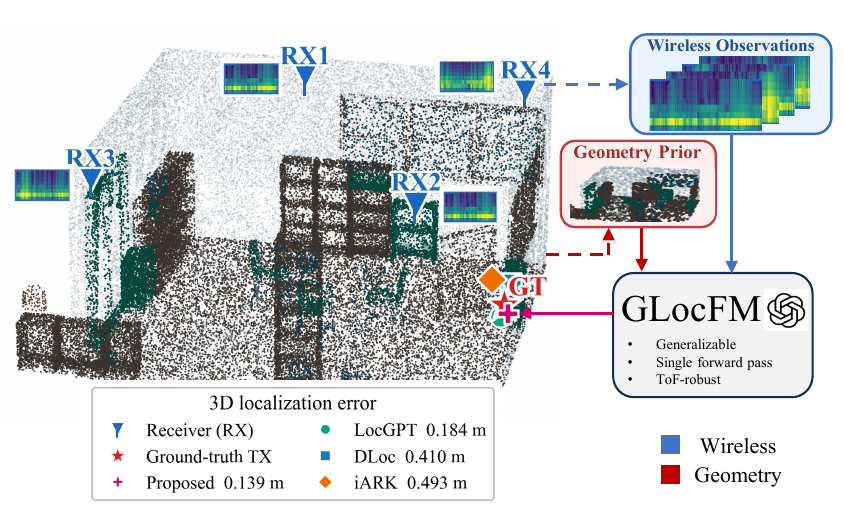}{}
    }
    \caption{The overview of the proposed \method{} which takes both geometry and wireless observation as input to produce 3D position of the transmitter.}
    \label{fig:typical-localization-main}
    \label{fig:architecture}
\end{figure}

Learning-based approaches, with their strong ability to capture important statistics and propagation features, have become mainstream methods for the complicated indoor localization problem. Existing learning-based wireless localization methods can be divided into two categories. The first maps channel state information (CSI) directly to coordinates without explicitly exploiting spatial information
\cite{zhao2024locgpt,an2020iark,ayyalasomayajula2020dloc,kotaru2015spotfi}. Such models offer efficient inference, but can only {learn correlations between wireless observations and Tx positions~\cite{studer2018channelcharting}} by exploring scene geometry implicitly, leading to limited generalization ability.
The second group explicitly incorporates spatial information, including 2D floor plans, 3D maps, and meshes, into wireless localizers \cite{vuckovic2021mapcsi,kanhere2025mapassisted,han2025rayloc,hehn2025wigatr,chu2026sigmap}. Spatial information can improve localization accuracy, but it also increases data-acquisition and modeling complexities: {spatial representations of scenes must be accurately constructed~\cite{ayyalasomayajula2020locap,suga2023indoor}}; more advanced neural models capable of processing images and 3D point clouds must be employed; and, in some methods, ray tracing \cite{han2025rayloc} and inference-time optimization \cite{hehn2025wigatr} must be performed for individual samples, leading to high latency. %Simplifying or globally pooling the scene reduces these costs but discards candidate-dependent propagation structure.
To this end, we aim to advance prior work by proposing a geometry-aware foundation model that can achieve superior localization performance and is capable of generalizing to unseen scenes using only one feed-forward pass.

A critical requirement for obtaining such a foundation model is high-quality training data. Unlike large language models (LLMs), which have access to vast amounts of data on the internet, wireless localization foundation models have far too little real-measurement data available for training. Public available datasets are given in \cite{torressospedra2014ujiindoorloc,alkhateeb2019deepmimo}, yet the former does not include 3D geometry, and the latter contains too few indoor scenes to meet the generalization objective. We therefore construct a multimodal dataset of 221 diverse geometric layouts, using Infinigen Indoors \cite{raistrick2024infinigenindoors} to produce furnished 3D indoor environments represented as 3D meshes, and use the widely adopted Sionna RT \cite{hoydis2023sionna}  to generate wireless signals within each indoor scenario.
%We retain registered point clouds, complex channel frequency responses (CFRs), calibrated receiver poses, and transmitter locations, and use a scene-disjoint split to train an efficient cross-scene localizer.

Given both wireless observations and scene geometry, we formulate indoor localization as an ML problem in which we first partition the space uniformly into a grid and evaluate which grid center maximizes the likelihood of the wireless observations conditioned on the geometry. Given that the exact likelihood function is intractable, we instead adopt a learned scoring function to model it. In particular, we first calculate the cosine similarity between the observed delay--angular spectrum and the spectrum predicted by considering the LOS and one-bounce reflection paths between each grid center and receiver. This cosine value serves as a prior for the subsequent learning-based scoring function, which directly predicts the probability that the Tx lies within each grid cell. A off-grid component prediction network ensures that the model outputs a continuous localization position rather than a discrete one. We also consider the scenario where an unknown ToF offset exists between the transmitter and receivers. To tackle this, we introduce arbitary ToF offsets during training such that the proposed \method{} is capable of extracting features that remain informative under a non-zero ToF offset.
Numerical experiments are performed to confirm the superiority of the proposed \method{} over the baselines. We further provide ablation studies to validate its effectiveness and robustness. 
Our contributions are:
\begin{itemize}
    \item We propose GLocFM, a geometry-aware foundation model grounded in the maximum-likelihood principle, enabling robust, cross-scene generalizable, 3D localization.
    \item We construct a multimodal wireless localization dataset comprising of 221 scenes with 3D geometry and wireless signals simulated via Sionna RT.
    \item We utilize the geometry prior to generate the delay-AoA spectrum of each grid center, and employ learned scoring function to calculate the likelihood of each grid. A ToF-robust GLocFM model is developped to handle unknown ToF offsets between the transmitter and receiver.
    \item We demonstrate mean-error reductions of 49.5--56\% on unseen synthetic scenes and 48.8\% on NeRF$^{2}$ dataset with real measurement. Ablation studies are performed to justify its effectiveness and robustness.
\end{itemize}

\section{Related Work}

\subsection{Localization without Geometry}
Most wireless localization systems {infer position directly from radio
measurements~\cite{xiong2013arraytrack,vasisht2016chronos,xiong2015tonetrack,xie2019mdtrack}} without explicitly modeling the environment. SpotFi
\cite{kotaru2015spotfi} jointly estimates AoA and ToF and selects the path most
likely to be LoS, but this decision is made purely in the signal domain. IARK
\cite{an2020iark} improves multipath-robust AoA estimation across IoT protocols, but introduces neither ranging nor geometric constraints.
The authors of LocGPT \cite{zhao2024locgpt} pretrain Transformers for direction estimation and
triangulation, while DLoc \cite{ayyalasomayajula2020dloc} fuses the AoA--ToF maps from multiple receivers into a location heatmap for localization that is also robust to ToF offsets.
Although efficient, these methods {learn the radio-to-location mapping~\cite{wang2015deepfi,wang2016phasefi,chen2017confi,zhang2022tips}} without checking how the observed paths could propagate through the scene. 
%Under blockage and multipath, similar angle--delay patterns may correspond to different locations, causing the learned representation to entangle localization cues with the training environment. Geometry provides an additional test: whether a candidate location admits physically plausible LoS and NLoS paths.
\subsection{Wireless Localization with Geometry}
Recently, researchers have started to incorporate scene geometry into localization.
MAP-CSI \cite{vuckovic2021mapcsi} backtracks dominant AoD--ToA components
through a map, while MAP-AT \cite{kanhere2025mapassisted} traces resolved
angle--delay paths in measured mmWave/THz environments. These approaches {make
the propagation geometry explicit~\cite{gentner2016channelslam}}, but rely on resolving a small number of
individual paths, which is difficult with an 80-MHz frequency band, and are mainly
demonstrated in 2D or single-site settings.
Another line of work utilize learning-based approach. LocUNet \cite{yapar2022locunet} combines city and path-loss maps for 2D urban localization, while GEAL \cite{keum2026geal} identifies reflection points corresponding to NLoS paths using 2D geometric information. Neither is designed for 3D indoor localization. RayLoc \cite{han2025rayloc} and Wi-GATr \cite{hehn2025wigatr} employ richer physical models, but localize by inverting a forward simulator through gradient optimization, requiring calibrated geometry, material information, or iterative inference.
SigMap \cite{chu2026sigmap} is more closely related to the proposed scheme because it also uses complex CFRs and treats a 3D map as a prompt to its foundation model. However, it focuses on outdoor scenarios evaluated on the DeepMIMO dataset, leaving the more challenging indoor localization problem  with more NLoS paths unresolved.

\section{GLocFM Dataset Generation}\label{sec:dataset_generation}
The proposed multimodal GLocFM dataset is comprised of 221 indoor scenes, whose geometric part is generated using Infinigen Indoors \cite{raistrick2024infinigenindoors}, and  the wireless part generated using Sionna RT. The scenes cover diverse furnished layouts including offices, bedrooms, bathrooms and kitchens. These scenes occupy areas from
$3.25\,\mathrm{m}^{2}$ to $63.51\,\mathrm{m}^{2}$, with a mean of $22.37\,\mathrm{m}^{2}$. Each scene consists of 3D object meshes 
%whose geometry and metadata specify the pose and optical material properties of each object.}
and before simulating the wireless channel in each scene, we assign physically plausible EM properties  to each mesh object according to its semantic type. Sionna RT then generates the channel response at each receiver antenna array.
%We further assume that the transmitter radiates electromagnetic waves isotropically, while the receivers are modeled as half-wavelength-spaced $4\times4$ uniform planar arrays (UPAs) where each attenna element adopts the 3GPP TR 38.901 antenna radiation pattern.
{Although Sionna RT operates on complete 3D meshes, such geometry is seldom available in practical deployments. %The 3D point cloud, on the other hand, can be collected more readily. 
Thus, we construct the GLocFM dataset by representing the scene geometry using a point cloud, obtained by sampling each mesh surface in proportion to its area. The transmitter and receiver positions used in the dataset are sampled from the free space, denoted by $\Omega_{\rm free}$, which is the volume not occupied by any object mesh.}
To be precise, let $\mathcal{S}$ denote the scene and let $\Omega\subset\R^3$ be its admissible interior volume. For object mesh $\mathcal M_j$, let $\mathcal O_j\subset\Omega, j \in [1, J]$ denote the closed region occupied by that object, we define $\Omega_{\rm occ}=\bigcup_{j=1}^{J}\mathcal O_j$ and $\Omega_{\rm free}=\Omega\setminus\Omega_{\rm occ}$.
{
%For each scene, we sample 20 transmitter positions and 20 receiver positions. 
To model ceiling-mounted infrastructure receivers and
mobile transmitters below them, receiver candidates lie within 0.15 m of the
ceiling, while transmitter heights span 10--70\% of the scene height. We split the 221 scenes into 177
training, 22 validation, and 22 test scenes.}

{The channel simulation uses a carrier frequency of 3.5 GHz, 80 MHz
bandwidth, and 1,024 OFDM subcarriers ($\Delta f=78.125$ kHz). We include specular and diffuse reflection with a maximum interaction depth of three, and model the transmit antenna pattern of the transmitter as isotropic.}
Each receiver is a half-wavelength-spaced $4\times4$ UPA adopting the 3GPP TR 38.901 antenna radiation pattern.
Let $\vect{Q}_r\in\mathrm{R}^{3\times3}$ denote the orientation of the $r$-th receiver with $r\in[1,N_R]$, which maps UPA-local coordinates to the world coordinates, and $\vect{Q}_r^\top$ performs the inverse mapping. 
%Under the Euler-angle convention used in our simulator, the angles $[0,\pi/2,0]$ map the local $+x$ boresight to world $-z$. Thus, $\vect{Q}_r^\top$ maps each world-frame path direction into the receiver-array frame.
{For every propagation path $p$ between the transmitter and
receiver element $a$, Sionna RT returns a complex path gain $\alpha_{p,a}$,
delay $\tau_p$, and incidence angle $\theta_p$. The receiver CFR can be expressed as:}
\begin{equation}
H_{r,a,k}=\sum_{p\in\mathcal{P}_{r}}
\alpha_{p,a}\exp(-j2\pi f_k\tau_p),
\end{equation}
{where $a \in [1, A]$ and $k \in [1, K]$ denote the antenna and subcarrier indices, respectively, and $\mathcal{P}_{r}$ is the set of paths connecting the
transmitter to the $r$-th receiver.}

%Of the $221\times20\times20=88{,}400$ links, 78,250 contain a direct component. The remaining 10,150 links (11.48\%) are NLoS-only. 
%This fraction comes from the traced geometry rather than a prescribed blockage probability; links with LoS may also contain strong reflections. 

\begin{figure*}[t]
    \centering
    \includegraphics[width=0.95\linewidth]{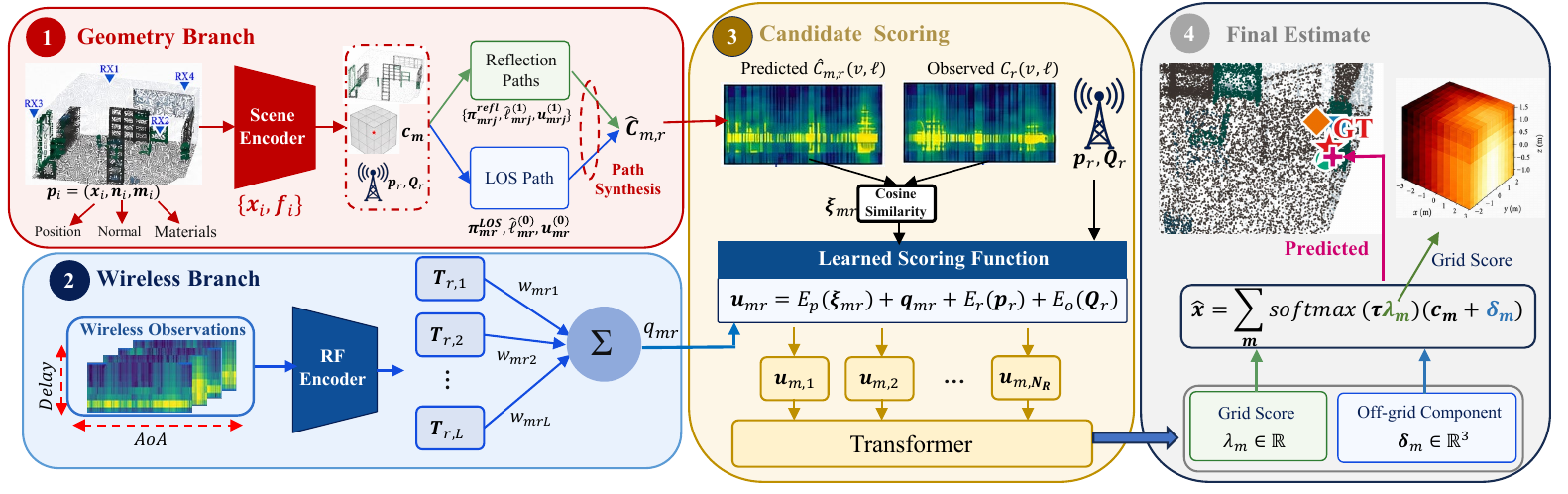}
    \caption{The flowchart of the proposed \method{} model.}
    \label{fig:glocfm_framework}
\end{figure*}

\section{GLocFM Model}
The localization problem can be formulated as an ML estimation problem for the transmitter position given the scene geometry, $\mathcal{S}$, and the observed CFRs at $\nrx$ receivers.
Given that the transmitter location is sampled from the free space, $\vect{x}\in \Omega_{\rm free}$, its  ML estimate can be expressed as:
\begin{equation}
\hat{\vect{x}}_{\rm ML}
=\arg\max_{\vect{x}\in\Omega_{\rm free}}
\log p(\{\vect{Y}_r\}_{r\in[1, \nrx]}\mid\vect{x},\mathcal S),
\label{eq:continuous-ml}
\end{equation}
where $\nrx$ denotes the number of receivers, and $\vect Y_r$ is the CFR at receiver $r$. Because this continuous search is impractical, we instead adopt a grid search together with a learned scoring function for each grid cell, as described below.

\subsection{Utilizing Geometric Priors}
\subsubsection{Generation of the Candidate Grid and Propagation Paths}
We first illustrate how we partition $\Omega_{\rm free}$ into a discrete set of candidate transmitter (Tx) positions and generate the corresponding propagation paths for each candidate Tx position.
We denote the position and orientation of the $r$-th receiver by $\vect{p}_r$ and $\vect{Q}_r$, respectively. We partition the space uniformly, creating $N_x=N_y=N$ horizontal cells and $N_z$ vertical cells, giving $M\triangleq N^2N_z$ candidates $\{\vect{c}_m\}_{m=1}^{M}$. 

For each candidate $\vect{c}_m$, we generate the corresponding propagation paths to each receiver $r$. Given that the LOS and the one-bounce reflection paths convey the majority of the transmit power, we simplify the propagation modeling of the proposed GLocFM framework by ignoring higher-order interactions. To start with, the LOS path connecting $\vect{c}_m$ and $\vect{p}_r$ has length
$d^{(0)}_{mr}=\|\vect{c}_m-\vect{p}_r\|_2$, delay $\tau^{(0)}_{mr}=d^{(0)}_{mr}/c$, and a direction $\vect{u}^{(0)}_{mr}=\vect{Q}_r^{\mathsf T}
\frac{\vect{c}_m-\vect{p}_r}{d^{(0)}_{mr}},$
expressed in the receiver's local coordinate system. 

We then consider one-bounce reflection paths. Let $\vect{a}_j$ be a potential reflection point, which lies on the surface of an object in the scene. It defines a one-bounce propagation path from candidate $\vect{c}_m$ to receiver $\vect{p}_r$:
\begin{align}
d^{(1)}_{mrj}&=\|\vect{c}_m-\vect{a}_j\|_2+
                 \|\vect{a}_j-\vect{p}_r\|_2,\\
\vect{u}^{(1)}_{mrj}&=\vect{Q}_r^{\mathsf T}
\frac{\vect{a}_j-\vect{p}_r}{\|\vect{a}_j-\vect{p}_r\|_2},
\end{align}
where the superscript indicates a single interaction.
These quantities are used to calculate key parameters such as amplitude, delay, and AoA, which contribute to the likelihood of each candidate Tx position, as detailed later.
%The anchors are latent alternatives rather than labeled ray interactions. The grid supplies path hypotheses, and an offset head recovers displacement within a local cell. We choose $N_z$ independently for the valid transmitter height range. Increasing $N$ improves horizontal resolution at a quadratic cost in receiver--candidate path queries.

%All angles are evaluated in the local receiver frame. Ceiling arrays can view the same world point from similar global directions but different local array coordinates. The stored Sionna orientation therefore enters every path direction used by the scorer.

\subsubsection{Scene Encoding}
We then illustrate how we extract essential wireless propagation features from the scene geometry. 
%Since we only consider the LOS and one-bounce reflection paths, we focus on  whether the LOS path is obstructed and the likelihood of the one-bounce reflection path to exist, which is detailed as follows.
As shown in Fig. \ref{fig:architecture}, we represent the scene as a 3D point cloud with associated features, which are fed into a scene encoder comprising multiple downsampling and self-attention layers that hierarchically reduce the number of points and aggregate features. In particular, the scene encoder first maps the coordinate $\vect{x}_i$, normal $\vect{n}_i$, and EM attributes $\vect{m}_i=[\varepsilon_{r,i},\sigma_i,S_i,K_{x,i}]$, comprising relative permittivity, electrical conductivity, diffuse-scattering coefficient, and cross-polarization coefficient, to a high-dimensional feature: $\vect{f}^{(0)}_i=\phi_0([\vect{x}_i,\vect{n}_i,\vect{m}_i]).$
Then, at hierarchy level $q, q\in [1, Q]$, farthest-point sampling (FPS) selects the subset $\mathcal I_q$, and, for each point $i\in\mathcal I_q$, a $k$NN network aggregates features from its $k$ nearest neighbors $\mathcal N_k(i)$:
\begin{align}
\vect{f}^{(q)}_i=\max_{j\in\mathcal N_k(i)}
\phi_q\!\left(\vect{x}^{(q-1)}_j-\vect{x}^{(q-1)}_i,
\vect{f}^{(q-1)}_j\right),
\label{eq:fps-point-feature}
\end{align}
where $\phi_q$ is an MLP layer and max denotes max-pooling operation. 
At hierarchy level $Q$, we apply a self-attention module to the features $\vect{f}^{(Q)}_i$ of the surviving points $\{\vect{p}_i,\vect{f}^{(Q)}_i\}_{i\in\mathcal I_Q}$ to obtain $\{\vect{f}_i\}_{i\in\mathcal I_Q}$, which are used to compute the geometric priors for each candidate Tx--receiver pair.
\subsubsection{Modeling Attenuation of Propagation Paths}
We first model the attenuation of a LoS path caused by obstruction and
free-space path loss. To represent obstruction, we construct the opacity field following the principle of convolutional occupancy networks \cite{peng2020convolutionaloccupancynetworks} using the surviving scene points $\{\vect{p}_i,\vect{f}_i\}_{i\in\mathcal I_Q}$. The resulting opacity field can be queried at an arbitrary 3D location by trilinear interpolation.
%Because the voxel field aggregates contributions from multiple anchors, its queried value is treated as a nonnegative learned opacity proxy rather than a normalized occupancy probability.
For a segment with endpoints $(\vect{x},\vect{y})$, we use $T$ uniformly
spaced interior query points, $\vect{s}_t = \vect{x} + \frac{t}{T+1} (\vect{y}-\vect{x}), t \in [1,T].$
Let $\rho_t\geq 0$ denote the opacity queried at $\vect{s}_t$, the accumulated opacity along the segment can be expressed as:
\begin{equation}
\mathcal O(\vect{x},\vect{y})
=
\frac{\|\vect{y}-\vect{x}\|_2}{T}
\sum_{t=1}^{T}\rho_t.
\end{equation}
We introduce soft visibility, $V(\vect{x},\vect{y})$, to quantify the obstruction level, which is defined as:
\begin{equation}
\begin{aligned}
\log V(\vect{x},\vect{y})=-\operatorname{softplus}(\gamma) \mathcal O(\vect{x},\vect{y}),
\end{aligned}
\label{eq:path-visibility}
\end{equation}
where $\operatorname{softplus}(\gamma)>0$ is a learned attenuation factor.
Consequently, $V(\vect{x},\vect{y})\in(0,1]$: it approaches one for a clear
segment and decreases along an obstructed segment.
The use of a learned soft visibility, rather than a binary ray-tracing label, also provides smoother gradients during training. The LOS geometric prior is then written as:
\begin{equation}
\begin{aligned}
\pi^{\rm LOS}_{mr}={}&
-\beta_{\rm L}\log(d^{(0)}_{mr}/d_0)
+\log V(\vect c_m,\vect p_r)\\
&+g_{\rm L}\!\left(d^{(0)}_{mr},
\log V(\vect c_m,\vect p_r)\right),
\end{aligned}
\label{eq:LOS-prior}
\end{equation}
where $\beta_{\rm L}$ is a learned scale factor initialized to the free-space path-loss exponent, $d_0$ is a positive constant to scale the distance, and $g_{\rm L}(\cdot)$ is a learned function parameterized by an MLP for calibration.

The geometric prior for a one-bounce path is evaluated along two segments
joined at a reflection point. Specifically, \eqref{eq:path-visibility}
models the obstruction from candidate $\vect c_m$ to the
reflection point $\vect a_j \in\mathcal I_Q$, and from $\vect a_j$ to receiver
$\vect p_r$, while two log-distance terms model distance-dependent
propagation loss. The two directions pointing from the reflection point toward the
candidate and receiver are denoted by $\vect v_{mj}=\frac{\vect c_m-\vect a_j}{d_{mj}}$ and $\vect v_{rj}=\frac{\vect p_r-\vect a_j}{d_{rj}}$, respectively.
At reflection point $j$, an MLP models the reflection-induced attenuation, denoted as $\alpha^{\rm refl}_{mrj}$, which takes these two directions, the surface normal $\vect n_j$, and the feature $\vect f_j$ associated with the reflection point as input:
\begin{equation}
\vect z_{mrj}=\operatorname{concat}\!\left(
\vect v_{mj},\vect v_{rj},\vect n_j,\vect f_j\right),
\qquad
\alpha^{\rm refl}_{mrj}=g_r(\vect z_{mrj}).
\label{eq:reflection-mlp-input}
\end{equation}
The overall geometric prior for the reflection path is then
\begin{align}
\pi^{\rm refl}_{mrj}={}&-\beta_{\rm L}[
\log(d^{(1)}_{mj}/d_0)+\log(d^{(1)}_{rj}/d_0)]\nonumber\\
&+\log V(\vect{c}_m,\vect{a}_j)
+\log V(\vect{a}_j,\vect{p}_r)+\alpha^{\rm refl}_{mrj}.
\label{eq:reflection-prior}
\end{align}
%We note that instead of adopting a binary LOS decision, keep this compatibility soft is essential to avoid a totally wrong decision due to the sparsity and imperfect of the point-cloud input. 
%Reflection points are likewise latent alternatives: no Sionna ray is assigned to an anchor during training. The reflection-anchor experiment therefore tests the path model, not a supervised ray classifier.

\subsection{Candidate Scoring Function}
We then illustrate how we design the scoring function given the observed wireless signals and the predicted ones. Instead of directly matching the CFRs, which are high-dimensional and sensitive to the estimated path parameters, we transform them into delay--angular spectrum which is calculated below:

For the observed wireless signals, we first transform its CFR into the delay--array spectrum via IFFT:
\begin{equation}
h_{r,a,\ell}=\frac{1}{K}\sum_{k=0}^{K-1}
H_{r,a,k}e^{j2\pi k\ell/K}.
\label{eq:cfr-delay-transform}
\end{equation}
Let $\{\vect v_g\}_{g=1}^{G}$ denote the set of steering vectors at the receiver, and let $\ell\in\{0,\ldots,L-1\}$ denote the delay-bin index.\footnote{We set $L \ll K$ as only the top delay bins are occupied.} For the $r$-th receiver, its delay--angular spectrum is:
\begin{equation}
C_r(\vect{v},\ell)=\log({|\sum_a h_{r,a,\ell}e^{-j2\pi\vect{e}_a^\top\vect{v}}|^2+\epsilon}).
\label{eq:delay-angular-spectrum}
\end{equation}
Here, $\vect v\in\mathcal V$ is expressed in the receiver's local coordinate system. The denominator normalizes the directional response by the received energy at the same delay bin. 

A propagation path can be represented by its delay, AoA, and  attenuation, $\pi^{\rm LOS}_{mr}$ and $\pi^{\rm refl}_{mrj}$, as defined in \eqref{eq:LOS-prior} and \eqref{eq:reflection-prior}, respectively. The delay $d_{mr}/c$ may correspond to a fractional number of delay bins, denoted by $\widehat\ell=d/\Delta_{\rm rng}$, where $\Delta_{\rm rng}$ is the range resolution determined by the signal bandwidth; interpolation is therefore required. In principle, the ideal interpolation function for a bandlimited signal is the sinc function. However, its sidelobes produce unstable gradients during training, leading to suboptimal performance, so we instead adopt the Gaussian surrogate:
\begin{equation}
\kappa_\ell(\widehat\ell)=\frac{\exp[-(\ell-\widehat\ell)^2/(2\sigma_\ell^2)]}
{\sum_{v=0}^{L-1}\exp[-(v-\widehat\ell)^2/(2\sigma_\ell^2)]}.
\label{eq:path-delay-kernel}
\end{equation}
The AoA parameter follows standard processing: for array element $a$ at position $\vect e_a$, 
%in the receiver's local coordinate system, 
the array response for an EM wave arriving from direction $\vect u$ is $e^{j2\pi\vect e_a^\top\vect u}$.
%The scalar LOS and reflection priors synthesize the complex delay--array spectrum as
Then, the predicted delay--array spectrum can be expressed as:
\begin{equation}
\begin{aligned}
\widehat{h}_{m,r,a,\ell}={}&
e^{\pi^{\rm LOS}_{mr}}e^{j2\pi\vect e_a^\top\vect u^{(0)}_{mr}}
\kappa_\ell(\widehat\ell^{(0)}_{mr})\\
&+\sum_j e^{\pi^{\rm refl}_{mrj}}
e^{j2\pi\vect e_a^\top\vect u^{(1)}_{mrj}}
\kappa_\ell(\widehat\ell^{(1)}_{mrj}).
\end{aligned}
\label{eq:composite-delay-array-spectrum}
\end{equation}
%where each exponential is implemented as $\mathcal A(\pi)$; the unclamped notation keeps Eqn.~(\ref{eq:composite-delay-array-spectrum}) readable. Thus the scalar priors control path magnitude, while phase is supplied by the complex steering response in Eqn.~(\ref{eq:path-array-response}); no additional per-path complex phase is predicted. Direct and reflected components are summed as complex values before the normalized delay--angular spectrum is formed, so their spatial phases interact coherently.

For candidate $m$, the predicted spectrum $\widehat C_{m,r}(\vect v,\ell)$ is obtained from (\ref{eq:delay-angular-spectrum}) by replacing $h_{r,a,\ell}$ with $\widehat h_{m,r,a,\ell}$. Stacking the sampled values over direction and delay gives $\mathbf C_r=[C_r(\vect v_g,\ell)]_{g,\ell}\in\mathbb R^{G\times L}$ and $\widehat{\mathbf C}_{m,r}$. We then measure the similarity between the  observed and predicted one as
\begin{equation}
s_{mr}=
\frac{\langle\mathbf C_r,\widehat{\mathbf C}_{m,r}\rangle_F}
{\|\mathbf C_r\|_F\|\widehat{\mathbf C}_{m,r}\|_F+\epsilon},
\label{eq:delay-angular-cosine}
\end{equation}

\subsubsection{Learned Scoring Function for Enhancement}
The candidate scoring function based on cosine similarity in \eqref{eq:delay-angular-cosine} may not be capable of fully exploiting the rich information contained in the scene geometry and the observed signals. Thus, we introduce a learned scoring function  for enhancement.

%In particular, the observed and predicted delay--angular spectra enter the neural scorer through their cosine similarity $s_{mr}$ in Eq.~(\ref{eq:delay-angular-cosine}), while the received-power feature $P_r$, receiver position, and receiver orientation are provided as additional inputs.
%These features are concatenated to form a feature vector $\boldsymbol\xi_{mr}$ for each candidate--receiver pair:
To start with, the learned scoring function maps the similarity score $s_{mr}$ along with the absolute received power $P_r$ at the $r$-th receiver to a feature vector $\boldsymbol\xi_{mr}$ containing geometric information: $\boldsymbol\xi_{mr}=\text{MLP}(\big[s_{mr}, P_r\big]),$
where $P_r$ can be understood as a `confidence' level of the similarity score. 
Then, the RF encoder shown in Fig.~\ref{fig:glocfm_framework}, which comprises Transformer layers, takes the CFR as input and produces tokens $\{\vect T_{r,\ell}\}_{\ell \in [1, L]}$ for  the $L$ delay taps. We obtain the feature vector for the $m$-th candidate by performing weighted pooling on $\vect T_{r,\ell}$ along the delay axis:
\begin{equation}
\begin{aligned}
\vect q_{mr}=\sum_{\ell=0}^{L-1}w_{mr\ell}\vect T_{r,\ell},
\end{aligned}
\label{eq:main-rf-delay-pooling}
\end{equation}
where $w_{mr\ell}$ are learnable parameters.
%Thus every candidate reads a different delay neighborhood of the same receiver token sequence. 
%After performing the feature extraction from the cosine similarity score $s_{mr}$ and the RF encoder, the candidate--receiver representation is
By aggregating the aforementioned  information which are essential for localization, we represent the candidate--receiver feature as:
\begin{equation}
\begin{aligned}
\vect u_{mr}= E_{p}(\boldsymbol\xi_{mr})
+\vect q_{mr} +E_r(\vect p_r)+E_o(\vect Q_r),
\end{aligned}
\label{eq:main-candidate-rx-token}
\end{equation}
where $E_{p}$, $E_r$, and $E_o$ are MLPs that map the path features, receiver position, and orientation to a high-dimensional space. 
%The LOS and reflection scalar columns are zero-masked, and direct reflector context is disabled. 
A self-attention block takes $\{\vect u_{mr}\}_{r\in [1, N_R]}$ as input and outputs $\vect f_m$ for each candidate,
%The separately predicted delay posterior is not used: RF tokens enter through $\vect q_{mr}$, and the unpooled complex-delay tensor enters through Eqn.~(\ref{eq:delay-angular-cosine}). 
which is then used to predict the final score, $\lambda_m$, for the $m$-th gird and its off-grid component, $\vect\delta_m$:
\begin{equation}
\begin{aligned}
\lambda_m&=h_{\rm score}(\vect f_m),\qquad
\vect r_m=h_{\rm off}(\vect f_m),\\
\vect\delta_m&=\tfrac12(\boldsymbol\Delta)\otimes
\tanh(\vect r_m).
\end{aligned}
\label{eq:implemented-candidate-logit}
\end{equation}
where $\boldsymbol\Delta=(\Delta_x,\Delta_y,\Delta_z)$ contains the spacings between neighboring candidate centers, and $\otimes$ denotes the element-wise product, which ensures that $\vect{c}_m+\vect{\delta}_m$ still lies within the cell of the $m$-th grid. The $h_{\rm score}$ and $h_{\rm off}$ are parameterized by MLP.
%The delay--angular-spectrum match enters through $E_{\rm path}(\boldsymbol\xi_{mr})$ rather than as a hand-weighted addition to the logit. 
With temperature $\tau>0$, the estimate is
\begin{equation}
\hat{\vect{x}}=\sum_{m=1}^{M}\operatorname{softmax}_m(\tau\lambda_m)
(\vect{c}_m+\vect{\delta}_m).
\label{eq:position-output}
\end{equation}
We note that the softmax function and the off-grid component $\vect{\delta}_m$ are essential to achieve a resolution finer than $\tfrac12\sqrt{\Delta_x^2+\Delta_y^2+\Delta_z^2}$, otherwise \method{} would simply produce the grid center, $\vect{c}_m$ with the highest $\lambda_m$ value.
%For NLOS-only links, reflected contributions can support the composite spectrum when the direct term is weak. For strong LOS links, the direct term can dominate without discarding compatible reflected energy.

We train the model by minimizing the $\ell_2$-distance between the predicted position and the ground truth:
\begin{equation}
\mathcal L_{\rm train} = \left\|\widehat{\vect x}_i-\vect x_i\right\|_2^2.
\label{eq:training-loss}
\end{equation}

\begin{table*}[t]
\centering
\small
\setlength{\tabcolsep}{3.5pt}
\begin{tabular*}{\textwidth}{@{\extracolsep{\fill}}lcccccccc@{}}
\toprule
& \multicolumn{4}{c}{$\nrx=3$} & \multicolumn{4}{c}{$\nrx=4$}\\
\cmidrule(lr){2-5}\cmidrule(l){6-9}
Method & Mean $\downarrow$ & Median $\downarrow$ & RMSE $\downarrow$ & $\widehat F_e(0.2)$ $\uparrow$
& Mean $\downarrow$ & Median $\downarrow$ & RMSE $\downarrow$ & $\widehat F_e(0.2)$ $\uparrow$\\
\midrule
\method{} & \textbf{0.1746} & \textbf{0.1181} & \textbf{0.2533} & \textbf{74.8}\%
& \textbf{0.1429} & \textbf{0.1032} & \textbf{0.1924} & \textbf{82.2}\%\\
LocGPT delay-AoA & 0.3972 & 0.2701 & 0.5085 & 36.2\%
& 0.2827 & 0.2197 & 0.4057 & 45.2\%\\
LocGPT AoA-only & 0.4143 & 0.2781 & 0.5453 & 32.6\%
& 0.3072 & 0.2225 & 0.4292 & 44.5\%\\
iARK & 0.6241 & 0.4409 & 0.8232 & 20.6\%
& 0.5079 & 0.3579 & 0.6718 & 28.0\%\\
DLoc & 0.6075 & 0.5072 & 0.7214 & 8.9\%
& 0.4746 & 0.3679 & 0.5875 & 16.7\%\\
Wi-GATr (RSRP) & 1.0480 & 0.8869 & 1.2215 & 0.9\%
& 0.8682 & 0.7281 & 0.9827 & 5.0\%\\
\bottomrule
\end{tabular*}
\caption{3D localization error (m) on the synthetic GLocFM dataset.}
\label{tab:main-results}
\end{table*}

\subsection{ToF-robust \method{}}
The \method{} model described above assumes perfect synchronization between
the transmitter and receivers. We further consider a TDoA setting in which the
receivers are mutually synchronized but share an unknown ToF offset relative to
the transmitter. For training sample $i$, the common offset $b_i$ produces the
same frequency-domain phase ramp at every receiver: $\widetilde H_{i,r,a,k}=H_{i,r,a,k}
 e^{-j2\pi\nu_k b_i}, r=1,\ldots,\nrx,$
where $\nu_k$ is the frequency of the $k$-th subcarrier. Following
\cite{koivisto2017jointclock,yang2022softlocalization}, we draw one offset per
training sample as $b_i\sim\mathcal{N}(0,\sigma_b^2)$. %The same sampled offset is applied to all receivers of that sample before constructing the delay--AoA observations.
This bias augmentation exposes the learned scoring function to different ToF offsets whose  loss function during training can be expressed as:
\begin{equation}
\mathcal L_{\rm train} = \mathbb{E}_{b\sim \mathcal{N}(0, \sigma^2_b)}\left\|\widehat{\vect x}_i(b)-\vect x_i\right\|_2^2.
\label{eq:training-loss-tof}
\end{equation}
This differs from the original model as the ToF-robust model optimized using \eqref{eq:training-loss-tof} is encouraged to exploit features that remain informative under a non-zero ToF offset, which are essential to retain robustness against unknown ToF offset in real deployment. The network architecture and the corresponding training methodology remain unchanged. During inference, the ToF-robust \method{} uses a single forward pass without the need to estimate the exact ToF offset.

\section{Simulation Results}
\subsection{Experimental Settings}
\subsubsection{Datasets}
We first evaluate the proposed \method{} model on the synthetic GLocFM dataset detailed above, which consists of 221 generated indoor scenes with corresponding wireless signals.
%Each scene contains 20 transmitter and 20 receiver locations with complex $4\times4$-UPA CFRs at 1,024 subcarriers, so no geometry or endpoint from a held-out scene appears during training. 
We also evaluate on the NeRF$^{2}$ dataset~\cite{zhao2023nerf2}, which provides 6,123 angular spectrum, of which 4,898 and 1,225 samples are used for training and testing, respectively. In particular, the authors perform real-world wireless experiments in which a single-antenna Tx at different locations communicates with a fixed receiver equipped with a $4\times4$ UPA. 
%Its official training/test counts are ; 490 official-training samples are held out for validation and test is evaluated once.

\subsubsection{Baselines}
Four learning-based localization baselines are considered, including localizers with and without geometry information.
We first implement the method in~\cite{hehn2025wigatr}, which takes the received signal power (RSRP) and the 3D map as inputs and performs optimization during inference.
We also implement localizers without geometry information. In particular, LocGPT AoA-only follows the original \cite{zhao2024locgpt}, in which only the received AoA spectrum is fed to the Transformer-based model. We furtherimplement LocGPT delay-AoA, in which delay information is also considered. IARK~\cite{an2020iark} is a multipath-resistant CNN-based AoA neural localizer, while DLoc~\cite{ayyalasomayajula2020dloc} utilizes both delay and AoA information, organizing the wireless signals as multiple AoA--ToF images. All these baselines are trained and tested using the same setup as the proposed \method{}.
%SpotFi~\cite{kotaru2015spotfi} motivates classical AoA--ToF processing, but we do not report it because this benchmark does not reproduce its commodity-WiFi hardware calibration or direct-path selection protocol. These are controlled neural baselines, not claims of protocol-identical reproduction.

\subsubsection{Metrics and Implementation Details}
The proposed scheme is trained using Adam with a learning rate of $10^{-4}$ for 200 epochs and a cosine scheduler on a device with 8 NVIDIA A6000 GPUs. The default GLocFM model adopts a $12\times12\times4$
candidate grid, a scene-encoder output cardinality of $|\mathcal{I}_p| = 256$, $L = 20$ retained delay taps, and an angular resolution of $5^\circ$. We report mean, median, root mean-square error (RMSE), and the CDF of 3D localization error which are obtained by averaging over multiple trials with different seeds. For
sample error $e_i$ and threshold $\tau$, the CDF is defined as:
\begin{equation}
{\widehat F_e(\tau)=\frac{1}{n}\sum_{i=1}^{n}
\mathbf{1}\!\left[e_i\leq\tau\right].}
\label{eq:cumulative-distribution-function}
\end{equation}

\begin{figure*}[t]
    \centering
    \setlength{\tabcolsep}{0pt}
    \begin{tabular}{@{}c@{\hspace{0.006\textwidth}}c@{\hspace{0.006\textwidth}}c@{}}
        \includegraphics[height=0.145\textheight]{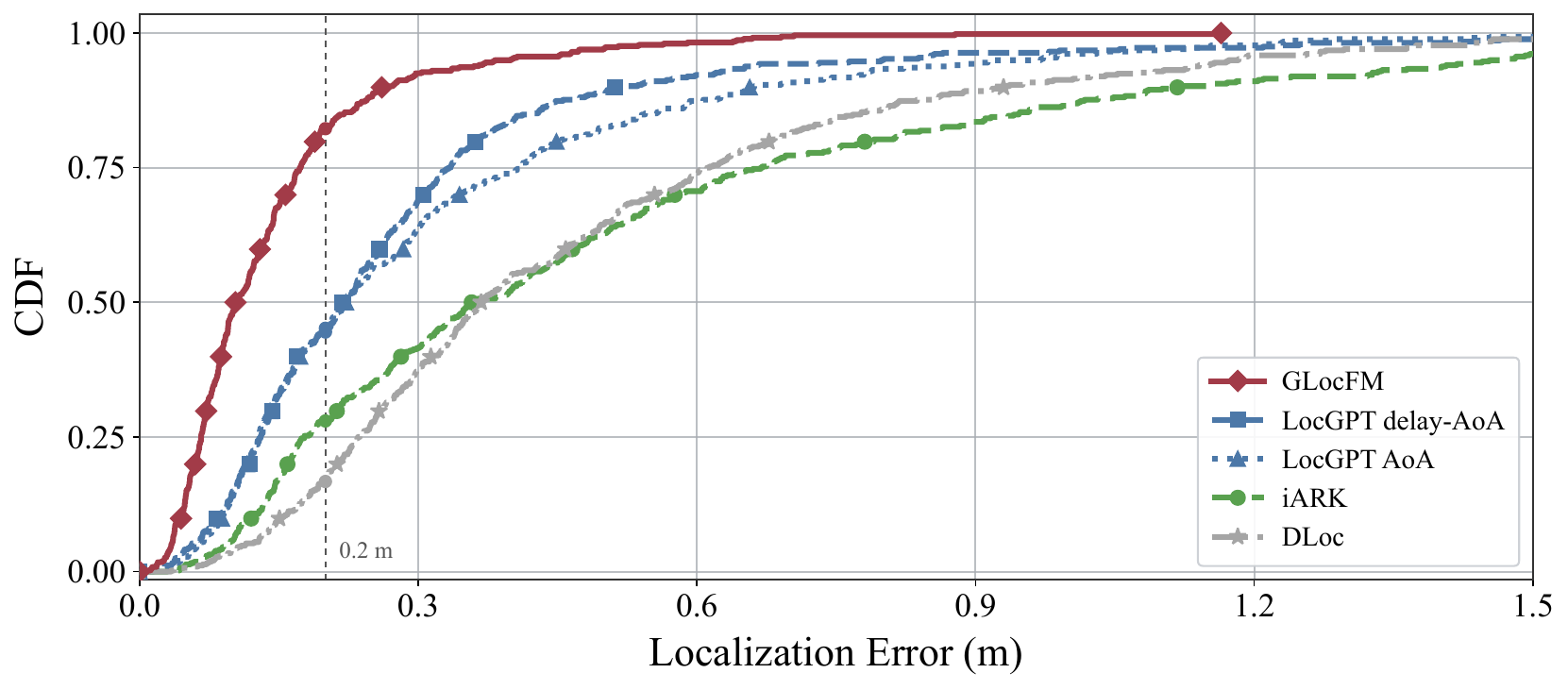} &
        \includegraphics[height=0.145\textheight]{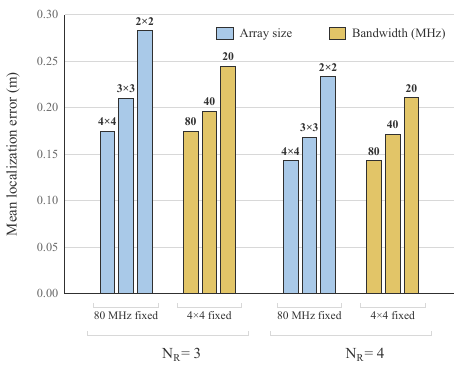} &
        \includegraphics[height=0.145\textheight]{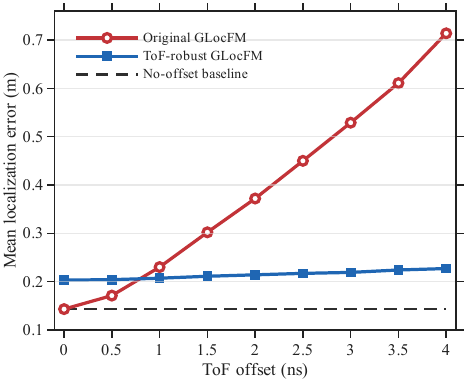} \\[-1pt]
        \textbf{(a)} & \textbf{(b)} & \textbf{(c)}
    \end{tabular}
    \caption{Evaluation of \method{}. \textbf{(a)}  CDFs achieved by different localization schemes. \textbf{(b)} Mean localization error with different system configurations. \textbf{(c)} Mean localization error under different ToF offsets for the original and ToF-robust \method{} models.}
    \label{fig:system-robustness-summary}
    \label{fig:synthetic-localization-results}
    \label{fig:localization-error-cdf}
\end{figure*}

\subsection{Comparison with Baselines}
\subsubsection{Comparison on the Synthetic GLocFM Dataset}
We first evaluate \method{} on the synthetic GLocFM dataset against the baselines described above. As shown in
Table~\ref{tab:main-results}, \method{} achieves the lowest mean, median, and
RMSE for different $N_R$ values.
In particular, it achieves a mean localization error of 0.1429 m 0.1746 m, which are 49.5\% and 56.0\% lower than that achieved by the strongest  LocGPT delay-AoA baseline for $\nrx=4$ and $\nrx=3$, respectively. We also note that the Wi-GATr  \cite{hehn2025wigatr} baseline which also adopt geometric information  achieves the worst localization performance. This is due to the fact that it is designed by only adopting the RSRP as input while ignoring the more informative delay--angular spectrum.

\begin{figure}[t]
    \centering
    \includegraphics[width=\columnwidth]{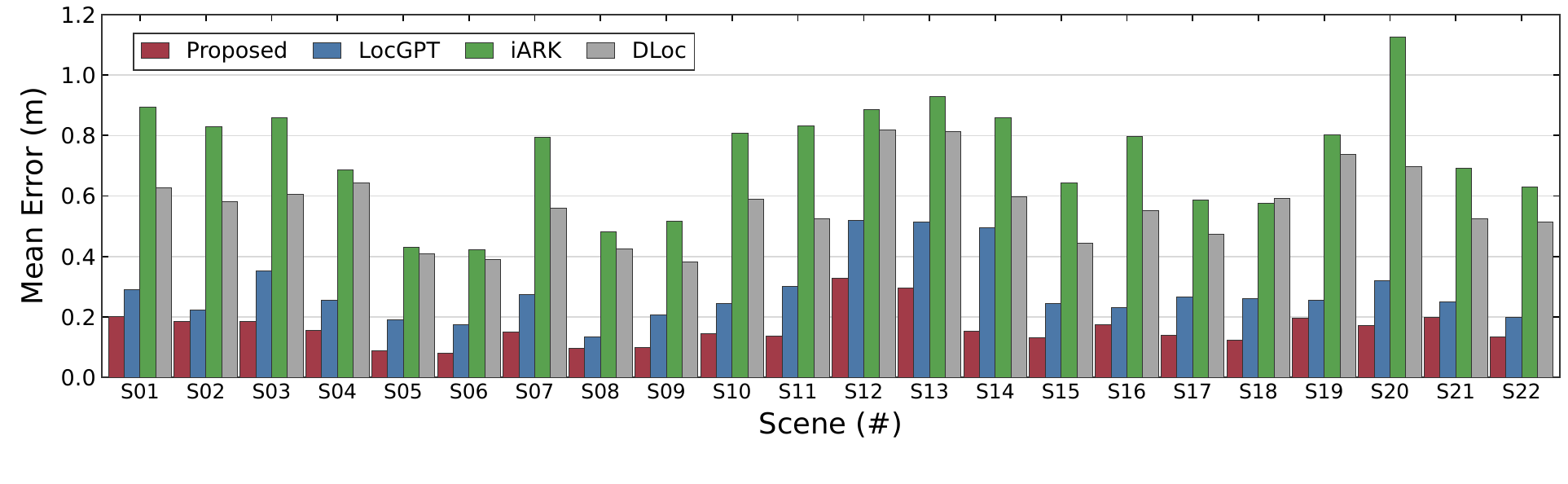}
    \caption{Mean localization error for each of the 22 scenes in the synthetic test dataset.}
    \label{fig:aligned-scene-error-main}
\end{figure}

It is interesting to observe that reducing the number of receivers to $N_R=3$ raises the mean error of \method{} only slightly, whereas the two LocGPT baselines show increases of approximately 0.1 m. This robustness is consistent with the reflected paths providing geometric constraints that complement direct-path triangulation. Fig.~\ref{fig:system-robustness-summary}(a) compares the empirical CDF for $\nrx=4$, where the proposed scheme achieves an accuracy of 82.2\%, surpassing the baseline schemes.
%At the 0.2-m threshold, the CDF values of \method{}, LocGPT delay-AoA, LocGPT AoA-only, iARK, and DLoc are 82.2\%, 44.5\%, 45.2\%, 28.0\%, and 16.7\%, respectively, exactly matching the corresponding entries in Table~\ref{tab:main-results}.
The larger CDF of \method{} throughout the low-error range indicates that its mean improvement is not driven solely by a few favorable samples.
Fig.~\ref{fig:aligned-scene-error-main} further reports the mean localization error for each of the 22 scenes in the test dataset, where \method{} consistently achieves lower errors than the baselines.

For a qualitative comparison, we visualize a representative localization sample in Fig.~\ref{fig:typical-localization-main}.
%, in which the ground-truth Tx lies near a wall in the scene, leading to a strong reflection path. 
It can be seen that \method{} achieves the best performance among all methods, whose localization  error is small relative to the scene scale.

\begin{table}[t]
\centering
\small
\setlength{\tabcolsep}{1.7pt}
\begin{tabular}{@{}lccccc@{}}
\toprule
Setting & Mean & Median & RMSE & $\widehat F_e(0.1)$ & $\widehat F_e(0.2)$ \\
\midrule
iARK & 0.2849 & 0.2666 & 0.3193 & 6.9\% & 29.7\% \\
LocGPT & 0.1244 & 0.0799 & 0.1816 & 63.5\% & 85.6\% \\

\midrule
\multicolumn{6}{@{}l}{\textit{Fine-tune using $p\%$ training data}} \\
\method{} ($p=100$) & \textbf{0.0637} & \textbf{0.0185} & \textbf{0.1429} & \textbf{86.6\%} & \textbf{91.4\%} \\
\method{} ($p=50$) & 0.0854 & 0.0273 & 0.1686 & 78.8\% & 87.1\% \\
\method{} ($p=25$) & 0.1304 & 0.0492 & 0.2233 & 67.3\% & 78.8\% \\
\bottomrule
\end{tabular}
\caption{Localization error on NeRF$^{2}$ dataset.}
\label{tab:nerf2-results}
\end{table}

\subsubsection{Evaluation on Real Measurements}
We next evaluate on the NeRF$^{2}$~\cite{zhao2023nerf2} dataset, in which each sample contains a 3D transmitter location and its corresponding angular power spectrum from the fixed receiver. Note that the authors did not provide CSI at the OFDM subcarriers; thus, no delay information is available. Consequently, we modify the original \method{} model
%must be modified. We retain the flowchart shown in Fig.~\ref{fig:architecture}, with
by applying minor changes to the predicted spectrum, $\hat{C}_r(\vect v,\ell)$, and the inputs to the RF encoder.

As shown in Table~\ref{tab:nerf2-results}, the modified \method{} model achieves a lower localization error than the baselines. Note that we train both the LocGPT and iARK baselines utilizing all samples in the training dataset. For the proposed scheme, however, we evaluate its generalizability by fine-tuning the model using $p$\% of the samples in the training dataset.
We observe that reducing the number of training samples from 100\% to 25\% raises mean localization error from 0.0637 to 0.1304 m and lowers the CDF at 0.1 m from 86.6\% to 67.3\%. It is worth mentioning that the weakened generalizability of the proposed scheme may be due to the differences in the experimental setup, such as antenna spacing and antenna patterns; investigating these effects is left for future work.

\subsection{Ablation Studies}
\subsubsection{Effects of Different GLocFM Designs}
Table~\ref{tab:path-grid-ablation} summarizes the impact of different components on the localization performance of the proposed \method{}. The ``Reference'' scheme corresponds to the default configuration with grid size $N = 12$.
\begin{table}[t]
    \centering
    \small
    \renewcommand{\arraystretch}{1.08}
    \setlength{\tabcolsep}{2pt}
    \begin{tabular}{@{}llrrrr@{}}
        \toprule
                Design & Variant & Mean & Median & RMSE & $\widehat F_e(0.2)$ \\
        \midrule
         {Reference} & --
          & 0.1429 & 0.1032 & 0.1924 & 82.2\% \\

        \midrule
        \multirow{2}{*}{Grid size}
          & $N=3$  & 0.4494 & 0.2825 & 0.7042 & 31.4\% \\
          & $N=16$ & \textbf{0.1360} & \textbf{0.1012} & \textbf{0.1830} & \textbf{84.2}\% \\
        \midrule
        %\addlinespace[2pt]
        
        %\addlinespace[2pt]
        Scorer
          & Cosine only & 0.3914 & 0.2977 & 0.5049 & 26.1\% \\
        \addlinespace[2pt]
        Path model
          & LoS only & 0.2285 & 0.1495 & 0.3515 & 66.0\% \\
        %Delay kernel & Sinc function & 0.1783 & 0.1269 & 0.2511 & 73.9 \\
        \bottomrule
    \end{tabular}
    \caption{Ablation study of different \method{} designs.}
    \label{tab:path-grid-ablation}
\end{table}

As can be seen, the size of the candidate grid is of vital importance: reducing it from $N=12$ to $N=3$ increases the mean error from 0.1429 m to 0.4494 m, whereas increasing it to $N=16$ yields an additional gain of 0.7 cm but nearly doubles the computational complexity.
The learned scoring function is essential, as removing it substantially worsens the error statistics, showing that simply adopting cosine similarity alone cannot resolve ambiguous candidates. We also demonstrate the effectiveness of the one bounce reflection path by showing a degradation of 7.5 cm when adopting the LoS-only \method{} model.
%Finally, a slight degradation is observed when the Sinc function is used to replace the Gaussian delay kernel.

\subsubsection{Different System Configurations}
We further perform an ablation study to evaluate the robustness of the proposed \method{} model with respect to different system bandwidths and UPA settings, as shown in Fig.~\ref{fig:system-robustness-summary}(b). Note that we fine-tune the model for only a single epoch when applying it to different configurations.
%Table~\ref{tab:input-ablation-main} places $\nrx=3$ and $\nrx=4$ side by side.
With $\nrx=4$, reducing the UPA from $4\times4$ to $3\times3$ and $2\times2$ increases the test error from 0.1429 m to 0.1684 m and 0.2332 m, respectively. The results obtained with $\nrx=3$ show the same trend. For both setups, reducing the bandwidth from 80 to 20 MHz increases the localization error by roughly 7 cm.
This is plausible because a smaller array weakens spatial resolution, while a lower bandwidth reduces range resolution and makes nearby ranges harder to separate. 
%Both types of resolution are critical to the learned scoring function, and weakening either one degrades localization performance.
%Reducing the number of delay bins is less damaging after adaptation because the measured bandwidth itself remains unchanged.

% === ARCHIVED TOF-OFFSET SIMULATION (posterior-marginalization wording) BEGIN ===
% \begin{table}[!t]
%     \centering
%     \scriptsize
%     \setlength{\tabcolsep}{2.2pt}
%     \caption{{Mean localization error (m) under a fixed positive
%     shared ToF offset.}}
%     \label{tab:clock-bias-main}
%     \resizebox{\columnwidth}{!}{%
%     \begin{tabular}{@{}l*{9}{c}@{}}
%     \toprule
%     ToF offset (ns) & 0 & 0.5 & 1 & 1.5 & 2 & 2.5 & 3 & 3.5 & 4\\
%     \midrule
%     {Synchronized \method{}} & 0.143 & 0.171 & 0.230 & 0.302 & 0.372
%     & 0.450 & 0.529 & 0.611 & 0.714\\
%     \midrule
%     {ToF-robust \method{}} & 0.203 & 0.204 & 0.207 & 0.211
%     & 0.214 & 0.217 & 0.219 & 0.224 & 0.227\\
%     \bottomrule
%     \end{tabular}}
% \end{table}
%
\subsection{Robustness to ToF Offset}
We then evaluate the localization performance of the proposed ToF-robust \method{} model and compare with the original one.
{The ToF-robust \method{} model is trained with $\sigma_b = 4$. Both schemes are evaluated with ToF offsets ranging from 0 to 4 ns.
As shown in Fig.~\ref{fig:system-robustness-summary}(c), the mean localization error of the original model increases dramatically from 0.1429 to 0.7141 m, demonstrating its vulnerability to ToF offset. The ToF-robust \method{} model performs worse under perfect synchronization, i.e., $b = 0$ ns. However, its mean error remains between 0.2044 and 0.2270 m over the same range of ToF offsets, outperforming the original model when $b\geq 1$ ns and reducing the localization error at 4 ns by 68.2\%. 
%These evaluations confirm the effectiveness of the proposed ToF-robust \method{} model.
}

% === ARCHIVED TOF-OFFSET SIMULATION (posterior-marginalization wording) END ===

\section{Conclusion}
In this paper, we proposed \method{}, a geometry-aware foundation model for 3D indoor wireless localization that jointly exploits WiFi measurements and 3D scene geometry. By matching observed and geometry-predicted delay--AoA spectra, \method{} effectively incorporates propagation priors for LoS and single-bounce reflection paths, while its ToF-robust variant handles imperfect synchronization. Experiments on synthetic and measured NeRF$^{2}$ datasets demonstrate substantial improvements over baselines, validating the effectiveness of scene geometry for accurate and generalizable wireless localization.

\bibliographystyle{plainnat}
\bibliography{references}
\end{document}